\documentclass[%
 reprint,
 amsmath,amssymb,
 aps,nofootinbib
]{revtex4-2}

\usepackage{graphicx}% Include figure files
\usepackage{dcolumn}% Align table columns on decimal point
\usepackage{bm}% bold math
\usepackage{hyperref}% add hypertext capabilities
\graphicspath{
	{operators/},{figures/}} 
\usepackage{subcaption}

\usepackage{xcolor}
\newcommand{\Tr}{\text{Tr}}
\newcommand{\im}{\mathrm{i}}
\newcommand{\qu}[1]{\overset{\circlearrowleft}{q}_{#1}}
\newcommand{\aqu}[1]{\overset{\circlearrowleft}{\bar{q}}_{#1}}
\newcommand{\quc}[1]{\overset{\circlearrowright}{q}_{#1}}
\newcommand{\aquc}[1]{\overset{\circlearrowright}{\bar{q}}_{#1}}
\newcommand{\spectraCaption}{lines show Eq. \eqref{eq:modified_nambu_goto} fitted to the simulation data and the width of the highlighted area around them corresponds to the error of the fit. $\lambda=0,\ 1,\ \ldots$ show the ground state, first excitation, and so on, depicted with the same color as the corresponding quantum number $N$. \textit{Fit for  $N:N_i\to N_f$} indicates the values of $N$ included in the fit. Gray highlighted areas show the charge separation $R$ included in the fit. }
\begin{document}

\preprint{APS/123-QED}

\title{Eight very excited spectra and one possible axion in SU(3) lattice gauge theory}% Force line breaks with \\
%\thanks{A footnote to the article title}%

\author{Alireza Sharifian}
\email{alireza.sharifian@tecnico.ulisboa.pt}
\author{Nuno Cardoso}
\email{nuno.cardoso@tecnico.ulisboa.pt}
\author{Pedro Bicudo}%
\email{bicudo@tecnico.ulisboa.pt}

\affiliation{%
CeFEMA, Departamento de F\'isica, Instituto Superior T\'ecnico (Universidade  de Lisboa),
Avenida Rovisco Pais, 1049-001 Lisboa, Portugal
}%

%\collaboration{CLEO Collaboration}%\noaffiliation

\date{\today}% It is always \today, today,
             %  but any date may be explicitly specified

\begin{abstract}
We compute the spectra of flux tubes formed between a static quark antiquark pair up to a significant number of excitations and for eight symmetries of the flux tubes, up to  $\Delta_u$, using pure  $SU(3)$ gauge lattice QCD in 3+1 dimensions. To accomplish this goal, we use a large set of appropriate operators, an anisotropic tadpole improved action, smearing techniques, and solve a generalized eigenvalue problem. Moreover, we compare our results with the Nambu-Goto string model to evaluate possible tensions which could be a signal for novel phenomena. Especially, we provide evidence for the coupling of a massive particle, say an axion, to the  $\Sigma_g^-$, $\Sigma_u^-$, and $\Sigma_u^{-*}$ flux tube with approximate masses $ 2.25\sqrt{\sigma}$, $ 1.85\sqrt{\sigma}$, $3.30\sqrt{\sigma}$, respectively.
\end{abstract}

%\keywords{Suggested keywords}%Use showkeys class option if keyword
                              %display desired
\maketitle

%\tableofcontents

\section{Introduction}

As gluons, force carriers of strong forces, have color charges, the gluonic fields are squeezed in the vacuum and form a flux tube. This is in contrast to the electromagnetic fields which spread out in the space. The dominant behavior of flux tubes are string-like. A confirmation for the string-like behavior is the Regge trajectories  \cite{Bicudo:2007wt,Bicudo:2009hm} observed in hadron spectra. The string theories also predict a linear potential between quarks which is confining and reproduces correctly \cite{Godfrey:1985xj,Isgur:1978xj} the confinement of quarks inside hadrons. 

Quantization of a relativistic string leads to a tower of excitations \cite{Alvarez:1981kc,Arvis:1983fp}, however different theoretical models exist for the excitations of hadrons, such as bag models for different sorts of hadrons \cite{DeTar:1983rw}, or a few-body potentials for mesons, baryons or hybrids \cite{Buisseret:2006wc}. Therefore, a first principle computation is important to test these models and search for novel phenomena. Numerous lattice QCD calculations \cite{Wilson:1974sk} have been devoted to study the excitations of the flux tube \cite{Juge:1999ie,Juge:2002br,Capitani:2018rox}. However, they only succeeded to compute a small number of excitations, up to two excitations for the most amenable symmetries of the flux tube. In this work, we continue our previous study of the $\Sigma_g^+$ spectrum \cite{Bicudo:2021tsc} and compute a significant number of excitations for other symmetries of the flux tube. We also compare our results with the Nambu-Goto \cite{Nambu:1978bd,Goto:1971ce} string model \cite{Aharony:2009gg}.

The Nambu-Goto string model is defined by the action
\begin{align}
S=-\sigma \int \mathrm{d}^2\Sigma, 
\label{eq:nambu_action}
\end{align}
where $\sigma$ is the string tension and $\Sigma$ is the surface of the worldsheet swept by the string. The energy of an open relativistic string with length $R$ and fixed ends is obtained as 
\begin{align}
&V(R)=\sqrt{\sigma^2 R^2+2\pi\sigma (N-(D-2)/24)},
\label{eq:nambu_goto}
\end{align}
where $N$ is the quantum number for string vibrations and $D$ is the dimension of space time. This expression is known as the Arvis potential \cite{Arvis:1983fp}.

For the excited states  the intrinsic width of the flux tube \cite{Cardoso:2013lla} should become negligible compared with the quantum vibrations of the string. The Nambu-Goto model should then be adequate to analyse the potentials we compute with lattice QCD. 
The agreement of the QCD flux tube and the Nambu-Goto is surprisingly good because Eq. \eqref{eq:nambu_action} is Lorentz invariant when $D=26$ while the QCD flux tube lives a four dimensional spacetime. Furthermore, while the ground state of Arvis potential ($N=0$) is tachyonic for small $ R\le \sqrt {\pi/(6 \sigma)}$, its value is imaginary, lattice QCD results are well-defined. It is even more interesting that the large $R$ expansion of Eq. \eqref{eq:nambu_goto} for the groundstate,  where the tachyon is replaced by the L\"uscher \cite{Luscher:1980iy} coulombic potential,
\begin{align}
V(R)=\sigma R-\frac{\pi}{12R}+\ldots, 
\end{align}
 is able to also fit correctly the lattice QCD potential for small and median $R$, matching at very short distances the correct potential matches perturbative QCD \cite{Karbstein:2014bsa}.
 However, the Lattice QCD results for some symmetries of the flux tube do not agree with the string model, even for large $R$ \cite{Dubovsky:2013gi,Athenodorou:2021vkw,Sharifian:2022proc}. The width of the flux tube \cite{Cardoso:2013lla} which is overlooked in this model or coupling of another particle to the string worldsheet \cite{Buisseret:2006wc} are important to understand these puzzles. 

 The hybrid mesons are actively searched in experiments.
	In the quark model, the gluonic degrees of freedom do not play any explicit role \cite{Godfrey:1985xj,Isgur:1978xj}. Consequently, when the flux tube is in its ground state,
	quantum numbers of conventional mesons $J^{PC}$ are determined by quarks angular momentum $L$ and spin $S$ where
	the parity $P=(-1)^{L+1}$ and the charge conjugation $C=(-1)^{L+S}$. These relations restrict the values of the quantum numbers of quark-antiquark pairs.
 Hence, other quantum numbers such as $J^{PC} = 0^{--} , 0^{+-} ,1^{-+} , 2^{+-},\ldots$ are not accessible in the quark model.  Particles with these quantum numbers are known as exotic mesons.  When the flux tube is excited, its angular momentum and spin contribute to  $L$ and $S$, therefore, exotic quantum numbers are obtained. The contribution of flux tube's angular momentum and its spin correspond to one gluon or more.   \cite{Olsen:2018_nsm,Meyer:2015HM}. Therefore, the spectrum of QCD is inevitably richer than that of the naive quark model.	Presently, there are some experimental \cite{Olsen:2018_nsm} candidates for exotic particles hybrids such as $\pi_1(1600)$  \cite{chung:2002}, $\pi_1(2000)$ \cite{LuM:2005} and $Y(4260)$ \cite{AlbertB:2005}. 

There is also an ongoing puzzle in the excited spectrum of mesons as reported \cite{Bugg:2004xu} in measurements by the Cristal Barrel detectors \cite{Aker:1992ny}: the Regge slope for radial excitations is similar to the one for angular excitations, $E^2 \propto (n_r+j)$ where $n_r$ and $j$  denote to the radial and rotation excitation, respectively. This cannot be explained with a quark model. In some string models, the Regge slope for radial excitations is two times larger than the slope for radial excitations $E^2 \propto(2n_r+j)$  \cite{Bicudo:2007wt}. 
Furthermore, a large degeneracy, larger than the chiral restoration symmetry \cite{Glozman:2007ek}, has been analysed \cite{Afonin:2006wt,Afonin:2007jd,Glozman:2012fj,Catillo:2018cyv}. Note that the chiral symmetry restoration predicts  the existence of the
approximately degenerate chiral partners of the high-lying well established states \cite{Glozman:2007ek}. Possibly there is a new principal quantum number \cite{Bicudo:2009hm} different from the one that is already present in the Nambu-Goto model.

There are two types of flux tubes, closed ones and open ones. The lattice counterparts of the closed flux tubes are the closed loops around the spatial torus. The spectrum of torelons is already studied in lattice QCD and the evidence for the existence of a new particle, say an axion, in the spectrum is reported  \cite{Dubovsky:2013gi,Athenodorou:2021vkw}. Besides, a model of four-dimensional relativistic strings with integrable dynamics on the worldsheet has been developed \cite{Dubovsky_2016}. In this work, we study open flux tubes whose lattice counterparts are Wilson loops.

\begin{figure}
	\includegraphics[scale=2]{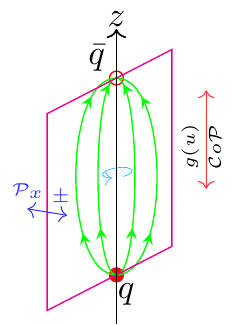}
	\caption{Symmetries of the flux tube.\label{fig:flux_tube_symmetries}}
\end{figure}

Classification of the open flux tube states is based on the three quantum numbers corresponding to the symmetries of the flux tube Fig. \eqref{fig:flux_tube_symmetries}.  The projection of angular momentum $J$ on the charge axis $J. \hat{R}$, where $\hat{R}$ is the unit vector along the charge axis,  denoted by $\Lambda$. It is common to use Greek letters $\Sigma$, $\Pi$, $\Delta$, $\Phi$, $\ldots$ to show $\Lambda=0,\ 1,\ 2,\ ,\ldots$, respectively. The second symmetry is the combination of charge conjugation and spatial inversion  about the midpoint between the quark and the antiquark $\mathcal{C}o\mathcal{P}$. Its eigenvalues $\eta_{CP}$  are $1(-1)$ and typically, are shown by $g(u)$. For $\Sigma$ states, there is an additional label $\epsilon$ which shows the eigenvalues of the reflection operator with respect to any plane containing the charge axis and is denoted by $+(-)$ for even (odd) states, respectively. Note that the energy of the gluons for $\Lambda\ge1$ states is unaffected by this reflection since such reflections only interchange the handedness of the state given by the sign of $J\cdot\hat{R}$. Consequently, the flux tube states are $\Sigma_g^+$, $\Sigma_g^-$, $\Sigma_u^+$, $\Sigma_u^-$, $\Pi_g$, $\Pi_u$, $\Delta_g$, $\Delta_u$, \ldots. We compute the spectra of these states up to a significant number of excitations.

\textcolor{black}{
We organize this paper into six sections. Sec. \ref{sec:string_states_cls} is devoted to the review of an effective string model to write the excitation of flux tubes in terms of right (left) circular polarization.  This helps us to identify the quantum number $N$ corresponding to each symmetry of the flux tube. In Sec. \ref{sec:lat_qcd}, we review the lattice QCD framework used in this work. Moreover,  we discuss how the appropriate operators for each symmetry are selected. We sieve the operators based on some criteria, and when the most appropriate  operator is found, we build up a tower of operators. Then, in Sec. \ref{sec:spectra},  we write down the explicit formula of the operators used for each symmetry to compute the Wilson correlation matrix. By solving the generalized eigenvalue problem, we find the spectrum of different symmetries of the flux tube. The spectra of flux tubes are compared with the Nambu-Goto model, and the departure from the Nambu-Goto model is quantified as well. In Sec. \ref{sec:axion_glueball}, we specifically analyze the $\Sigma_u^-$ and $\Sigma_g^-$ spectra whose spectra have a clear deviation from the Nambu-Goto string model. We show evidence for the existence of new particles in these spectra leading to a departure from the Nambu-Goto model. Finally, in Sec. \ref{sec:conclusion}, we conclude our work and discuss the outlook.
}
\section{Classification of an open string states \label{sec:string_states_cls} }

In this section, we classify the flux tube stats based on their quantum number. In the Nambu-Goto action, $d^2\Sigma$ is the surface element expanded as
\begin{align}
	\mathrm{d}^2\Sigma=\mathrm{d}^2 \xi\sqrt{g}, 
\end{align}
	where $g\equiv \mathrm{det}g_{\alpha\beta}$ and 
\begin{align}
	g_{\alpha\beta}=\partial_{\alpha}X_\mu \partial_\beta X^\mu
\end{align}
is the induced metric on the string world-sheet and $X^\mu$ are spacetime coordinates of the string. Furthermore, $\xi=(\xi^0, \xi^1)$ denotes to the parameters defining the world-sheet. This action is invariant under reparametrization, so to perform the calculations, we first select the so-called ``physical gauge''. 
In the physical gauge, world-sheet parameters are identified with longitudinal degrees of freedom of the string: $\xi^0=X^0$ and $\xi^1=X^1$. So, the string action only includes $(D-2)$ degree of freedom corresponding to the traverse displacement of the string from its equilibrium position. 
\begin{align}
	g =1&+\partial_{0}X_i\partial_{0}X^i+\partial_{1}X_i\partial_{1}X^i\nonumber\\
	&+\partial_{0}X_i \partial_{0}X^i \partial_{1}X_i \partial_{1}X^i - (\partial_{0}X_i \partial_{1}X^i)^2.
\end{align} 
The low energy expansion of the Nambu-Goto action for  some suitable redefinition of the fields  can be written as 
\begin{align}
	S=S_{cl}+\frac{\sigma}{2}\int \mathrm{d}^2 \xi \partial_\alpha X_i . \partial^\alpha X^i + \ldots, 
\end{align}
where $S_{cl}$ describes the usual perimeter-area term and 
the second term,  is the effective action proposed by L\"uscher,  M\"unster, and Weisz  as a effective string action \cite{Caselle:2021EST,Luscher:1980iy}. They achieved to find a correction for the linear confining potential between a quark-antiquark pair commonly known as the L\"uscher term, 
\begin{align}
V=c+\sigma R+\frac{\pi(D-2)}{24R}+\mathcal{O}(1/R^2).
\end{align}
The stationary states are found by expressing the traverse displacement fields $X^i$ in terms of normal modes. These modes have energies $m\omega$ for positive integer $m$ and $\omega=\pi/R$. We can define right $(+)$ and left $(-)$ circularly polarized ladder operators $a_{m_\pm}^\dagger$. The string eigenmodes are then,
\begin{align}
\prod_{m=1}^{\infty}	\left(( a_{m^+})^{n_{m_+}} 	( a_{m^-}^\dagger)^{n_{m_-}}\right)|0\rangle, 
\end{align}
where $|0\rangle$ indicates the ground state of the string, and $n_{m^+}$ and $n_{m^-}$ are the occupation numbers which take values $0, 1,2, \ldots$. We can obtain quantum number $N$, $\Lambda$ and $\eta_{CP}$ as the following:
\begin{align}
&N=\sum_{m=1}^{\infty} m (n_{m^+}+n_{m^-}),\\
&\Lambda=\big|\sum_{m=1}^\infty \left(n_{m^+}-n_{m^-}\right)\big|, \\
&\eta_{CP}=(-1)^N.
\end{align}
In this limit, the energy of the system is obtained as
\begin{align}
	E=E_0+\frac{N\pi}{R}.
\end{align}

In Table. \eqref{table:quantum_numbers}, we classify the flux tube states based on their energy. As we can see, there are multifold degeneracy for some states, for example, the energy level $N=1$ has two-fold degeneracy corresponding to $J.\hat{R}=\pm1$ \cite{Juge:2003ge}.
\begin{table}[h]
\begin{ruledtabular}
	\begin{tabular}{cccc}
		Excitation &
		Symmetry&
	State&\\
		\colrule
		$N=0$& $\Sigma_g^+$& $|0\rangle$& \\
		\colrule
		$N=1$& $\Pi_u$& $a^\dagger_{1^+}|0\rangle$& $a^\dagger_{1^-}|0\rangle$ \\
		\colrule
		$N=2$ & ${\Sigma_g^{+}}'$& $a^\dagger_{1^+} a^\dagger_{1^-}|0\rangle$&  \\
		& $\Pi_g$& $a^\dagger_{2^+}|0\rangle$& $a^\dagger_{2^-}|0\rangle$ \\
		& $\Delta_g$& $(a^\dagger_{1^+})^2|0\rangle$& $(a^\dagger_{1^-})^2|0\rangle$ \\
		\colrule
		$N=3$& $\Sigma_u^{+\prime}$& $(a^\dagger_{1^+}a^\dagger_{2^-}+a^\dagger_{1^-}a^\dagger_{2^+})|0\rangle$&  \\
	& $\Sigma_u^-$& $(a^\dagger_{1^+}a^\dagger_{2^-}-a^\dagger_{1^-}a^\dagger_{2^+})|0\rangle$&  \\
		& $\Pi_u'$& $a^\dagger_{3^+}|0\rangle$& $a^\dagger_{3^-}|0\rangle$ \\
		& $\Pi_u''$& $(a^\dagger_{1^+})^2 a^\dagger_{1^-}|0\rangle$&  $a^\dagger_{1^+} (a^\dagger_{1^-})^2|0\rangle$ \\
			& $\Delta_u$& $a^\dagger_{1^+} a^\dagger_{2^+}|0\rangle$&  $a^\dagger_{1^-} a^\dagger_{2^-}|0\rangle$\\
			&$\phi_u$& $(a_{1^+}^\dagger)^3|0\rangle$&  $(a_{1^-}^\dagger)^3|0\rangle$\\
		\colrule
		$N=4$ & ${\Sigma_g^{+}}''$& $a^\dagger_{2^+} a^\dagger_{2^-}|0\rangle$&  \\
	 & ${\Sigma_g^{+}}'''$& $(a^\dagger_{1^+})^2 (a^\dagger_{1^-})^2|0\rangle$&  \\
	  & ${\Sigma_g^{+(\mathrm{iv})}}$& $(a^\dagger_{1^+}a^\dagger_{3^-}+a^\dagger_{1^-}a^\dagger_{3^+})|0\rangle$&  \\
	   & ${\Sigma_g^{-}}$& $(a^\dagger_{1^+}a^\dagger_{3^-}-a^\dagger_{1^-}a^\dagger_{3^+})|0\rangle$  \\
	   &$\Pi_g'$&$a^\dagger_{4^+}|0\rangle$& $a^\dagger_{4^-} |0\rangle$\\
	   &$\Pi_g^{''}$&$(a^\dagger_{1^+})^2 a^{\dagger}_{2^-}|0\rangle$& $(a^\dagger_{1^-})^2 a^{\dagger}_{2^+}|0\rangle$\\
	   &$\Pi_g^{'''}$&$a^\dagger_{1^+} a^\dagger_{1^-} a^{\dagger}_{2^+}|0\rangle$& $a^\dagger_{1^+} a^\dagger_{1^-} a^{\dagger}_{2^-}|0\rangle$\\
	   
	   &$\Delta_g'$&$a^\dagger_{1^+}a^\dagger_{3^+}|0\rangle$&$a^\dagger_{1^-}a^\dagger_{3^-}|0\rangle$\\

	   &$\Delta_g^{''}$&$(a^\dagger_{2^+})^2|0\rangle$&$(a^\dagger_{2^-})^2|0\rangle$\\

	   &$\Delta_g^{'''}$&$(a^\dagger_{1^+})^3a^\dagger_{1^-}|0\rangle$&$a^\dagger_{1^+}(a^\dagger_{1^-})^3|0\rangle$\\
	   &$\Phi_g$&$(a^\dagger_{1^+})^2a^\dagger_{2^+}|0\rangle$&$(a^\dagger_{1^-})^2a^\dagger_{2^-}|0\rangle$\\
	   &$\Gamma_g$&$(a^\dagger_{1^+})^4|0\rangle$&$(a^\dagger_{1^-})^4|0\rangle$\\
	\end{tabular}
\end{ruledtabular}
\caption{Low-lying string states for a fixed ends open string \cite{Juge:2003ge}.\label{table:quantum_numbers}}
\end{table}

\section{Lattice QCD methodology\label{sec:lat_qcd}}
In this section,  we outline the lattice QCD framework  used in our calculations.

\subsection{Extracting hadron masses in lattice QCD}
\begin{figure}[ht]
	\includegraphics[scale=1.15]{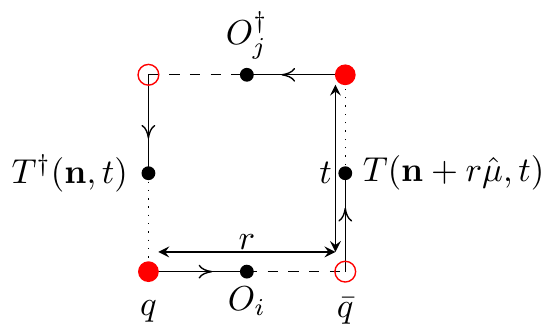}
	\caption{ A closed loop corresponds to the entry $\mathcal{C}_{i,j}(r,t)$ of the correlation Wilson matrix.\label{fig:wilson_loop}}
\end{figure}

To compute the spectra of the flux tube, we first compute the Wilson correlation matrix $\mathcal{C}(r, t)$. The entry $\mathcal{C}_{i,j}(r, t)$ of the Wilson correlation matrix is the expectation value of spatial-temporal closed loops, Fig. \eqref{fig:wilson_loop}, whose spatial sides are replaced with operators $O_i$ and $O_j$ having identical symmetry to the flux tube of interest. 
Afterwards, we find generalized eigenvalues $\lambda$ \cite{Blossier:2009kd,Dudek:2009qf,Dudek:2010wm,Bicudo:2021qxj} of the Wilson correlation matrix , 
\begin{align}
	\mathcal{C}(r, t)\vec{\nu}_n=\lambda_n(r, t) \mathcal{C}(r, t_0) \vec{\nu}_n, 
\end{align}
where we set $t_0=0$. Consequently, we obtain a set of time dependent eigenvalues $\lambda_n(t)$ for each $r$. Then, we order the eigenvalues and plot the effective mass defined as
\begin{align}\label{eq:effective_mass}
	E_i(r, t)=\ln \frac{ \lambda_i(r, t)}{\lambda_i(r, t+1)}.
\end{align}
The plateau in the effective mass plot corresponds to the energy $E_i(r)$. 

The generalized eigenvalue problem appears in our calculation due to the variational method techniques to find the closest state function to  the unknown physical state of the system. This method improves the overlap of the approximate state with the real state of the system as the number of operator bases increases. On the other hand, in a realistic calculation,  including more operator bases enhances the statistical noises which  affects the diagonalization of the correlation. Moreover, we have  a limited amount of available memory on our computers. Hence, we should choose a set of operators  which have a better overlap with the physical state of the system.

\subsection{Construction of operators \label{sec:operators_construction}}
As introduced  in the previous section, the entry of the Wilson correlation matrix $\mathcal{C}(r, t)$, is the correlation of an creation $O_i (\bold{n}, r\hat{\mu}, t_0)$ which create a pair of quark antiquark at time $t_0$ and the annihilation operator $O_j^\dag (\bold{n}, r\hat{\mu}, t_0+t)$  at time $t_0+t$ which annihilates them. So, to compute the Wilson correlation matrix, first we should construct the lattice operators with the desired quantum numbers. In the following, we outline how to build up the an operator with quantum numbers $\Lambda$, $\eta$, and $\epsilon$.

\subsubsection{Angular momentum, $\Lambda$}

Let us start with the continuum state and consider the trial state
whose  creation operator is
\begin{equation}\label{eq:trial}
	\underbrace{|\Psi_{\text{Hybrid}}\rangle_{S;\Lambda}}_{\text{trial state}}=\underbrace{\int_{0}^{2\pi}\mathrm{d}\phi\exp(\mathrm{i}\Lambda\phi)R(\phi)O_{S}}_{\text{creation operator}}|0\rangle,
\end{equation}
where $R(\phi)$ denotes a rotation by angle $\phi$ around the $z$-axis. Operator $O_{s}$ is an extended Wilson line
connecting a quark to an anti-quark. It can be written as 
\begin{equation}
	\hat{O}_{S}|0\rangle=\bar{Q}(-r/2)U_S(-r/2,r/2)Q(r/2)|0\rangle,
\end{equation}
where $Q(r/2)$ and $\bar{Q}(-r/2)$ are the creation operator of a \textit{spinless}
quark-antiquark pair. To show that the trial state of Eq. \eqref{eq:trial} has a definite
angular momentum $\Lambda$, we consider the effect of the rotation operator $\hat{R}(\alpha)$ on this state,
\begin{eqnarray}
	\hat{R}(\alpha)|\Psi_{\text{hybrid}}\rangle_{S;\Lambda} & =&\exp(-\mathrm{i}J_{z}\alpha)|\Psi_{\text{hybrid}}\rangle_{S;\Lambda}\nonumber\\
	&= & \exp(-\mathrm{i}J_{z}\alpha)\int_{0}^{2\pi}\mathrm{d}\phi\exp(\mathrm{i}\Lambda\phi)R(\phi)O_{S}|0\rangle\nonumber\\
	&= & \int_{0}^{2\pi}\mathrm{d}\phi\exp(\mathrm{i}\Lambda(\phi-\alpha)R(\phi)O_S|0\rangle\nonumber\\
&	=&\exp(-\mathrm{i\Lambda\alpha)|\Psi_{\text{hybrid}}\rangle_{S;\Lambda}}.
\end{eqnarray}
If we consider the expansion for the infinitesimal angle $\alpha$,
we obtain 
\begin{equation}
	J_{z}|\Psi_{\text{hybrid}}\rangle_{S;\Lambda}=\Lambda|\Psi_{\text{hybrid}}\rangle_{S;\Lambda},
\end{equation}
so, the trial state $|\Psi_{\text{hybrid}}\rangle_{S;\Lambda}$ is
a state with angular momentum $\Lambda$. The lattice version of
Eq. \eqref{eq:trial} is 
\begin{equation}
	|\Psi_{\text{hybrid}}\rangle_{S;\Lambda}=\sum_{k=0}^{3}\exp(\frac{i\pi\Lambda k}{2})R(\frac{\pi k}{2})O_{s}|0\rangle\label{eq:trial_lat}.
\end{equation}

\begin{figure}
	\centering
	\begin{minipage}[t]{1\columnwidth}%
		\centering
		\includegraphics{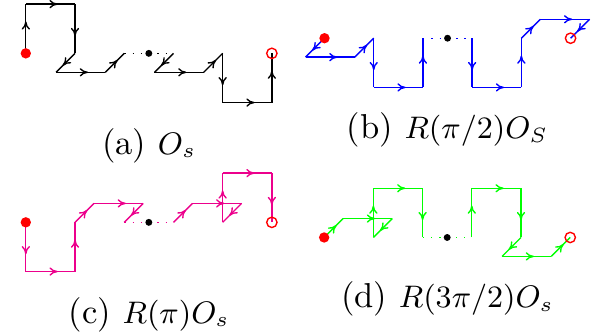}%
	\end{minipage}
	\caption{The effect of the rotation operator on an arbitrary operator $O_{s}$ is shown. The summation
		of all the components with appropriate phases $\exp(\mathrm{i}\Lambda\phi)$
		is the eigenstate of $J_{z}$ with eigenvalue $\Lambda$.\label{fig:rotation}}
\end{figure}

Since we are working with an ensemble of cubic lattices (note that it is cubic in space), the rotation angles are restricted
to multiples of $\pi/2$. It is clear from Eq. \eqref{eq:trial_lat} that 
the trial state for a $\Lambda$ is similar to the state $\Lambda'=\Lambda+4n$
where $n\in\{0,1,2,\ldots\}$. For example, in the $\Sigma_{g}^{+}$
channel, we may observe a signal for $\Gamma\ (\Lambda=4)$ channel too. In   Fig. \eqref{fig:rotation}, we show the effect of the rotation operator $R(k\pi/2)$ applied on an arbitrary extended Wilson line.

\subsubsection{Charge conjugation, parity, and Inversion}
\begin{figure}
	\begin{minipage}[t]{1\columnwidth}%
		\centering
		\includegraphics{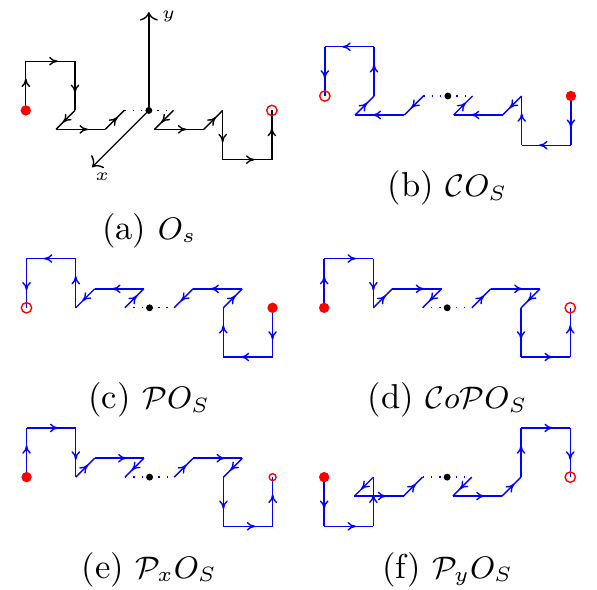}%
	\end{minipage}
	\caption{Effect of the charge parity $\mathcal{C}$, special reflection about the mid point  of the charge axis $\mathcal{P}$, and the inversion operator with respect to $x(y)$-plane $\mathcal{P}_{x(y)}$  on an arbitrary operator $O_S$. }\label{fig:Effect-of-some-operator}
\end{figure}

So far, we could construct the state with angular momentum $\Lambda$.
Although, it is possible that this state accidentally be the eigenstate
of $\mathcal{P}o\mathcal{C}$ or $\mathcal{P}_{x}$ for some specific
$U_{S}(-\mathbf{r}/2,\mathbf{r}/2)$, \iffalse for example, the operator $O(l,0)$ shown in  Fig. \eqref{fig:MO0} for $\Sigma_{g}^{+},\Pi_{u},\Delta_{g}$ channels,\fi it might be needed to add some extra terms to $|\Psi_{\text{hybrid}}\rangle_{S;\Lambda}$
to get $|\Psi_{\text{hybrid}}\rangle_{S;\Lambda_{\eta}^{\epsilon}}$.
We can see that the following changes happen when we apply $\mathcal{P}_{x}$
and $\mathcal{P}o\mathcal{C}$ on each component $O_{S}|0\rangle$ as  we illustrate them in Fig. \eqref{fig:Effect-of-some-operator} too:
\begin{enumerate}
	\item The charge conjugate operator changes the direction of each parallel
	transport $U_S(-r/2,r/2)$ or we can say the link is  replaced with  its conjugate
	transpose. Note that if there is a loop on the staple, they direction of the loop changes as well. Intuitively, we can say that charge conjugate replaces a source with a sink
	and vice versa, see Fig. (\ref{fig:Effect-of-some-operator}-b). 
	\item Spatial reflection  $\mathcal{P}$ with respect to the midpoint of the quark-antiquark separation changes
	the direction of each line as it reflects the links to the opposite
	side and also, it changes the location of the quark and anti-quark, see Fig. (\ref{fig:Effect-of-some-operator}-c). In Fig. (\ref{fig:Effect-of-some-operator}-d), the combination of the charge conjugation $\mathcal{C}$ and spatial reflection $\mathcal{P}$ was depicted.

	\item Inversion with respect to any plane containing the molecular axis operators. $\mathcal{P}_x$ or $\mathcal{P}_y$ is
	the mirror image of the operator with respect to $x$-plane or $y$-plane, notice charges are on $z$-axis, see Figs. (\ref{fig:Effect-of-some-operator}-e) and (\ref{fig:Effect-of-some-operator}-f). Moreover, there are two more planes we can define in a $4D$ lattice: a plane passes through the area $1$ and $3$ of the coordinate system, another passes through the area $2$ and $4$ of the coordinate system. A state has a well defined $\epsilon$, if it has the same value for $\epsilon$ under the reflection with respect to these 4 planes. 
\end{enumerate}
Considering these behaviours, to construct $|\Psi_{\text{hybrid}}\rangle_{S;\Lambda_{\eta}^{\epsilon}}$,
we should add some more components with the correct phase to Eq. \eqref{eq:trial_lat}. We use trial and error to construct a state with quantum number $\eta$ and $\epsilon$ of the interest. Some suboperators might not be appropriate for a symmetry, in this case, we end up with a null operator as we add more components to generate the state with the desired $\eta$ and $\epsilon$. 
 
\subsection{Action}
As the first step, we generate an ensemble of $SU(3)$ configurations using the anisotropic a tadpole improved action $S_{II}$ developed in Ref. \cite{Morningstar:1996ze},  
\begin{align}
	S_\text{II} =& \beta \big( \frac{1}{\xi}     \sum_{x,s>s'}  \big[\frac{5 W_{s,s'}}{3 u_s^4} - \frac{W_{ss,s'}+W_{s,s's'}}{12 u_s^6} \big]\nonumber\\
	&+  \xi \sum_{x,s} \big[\frac{4 W_{s,t}}{3 u_s^2 u_t^2} - \frac{W_{ss,t}}{12 u_s^4 u_t^2}  \big]\big), \label{eq:S2}
\end{align}
where $\beta=6/g^2$ is the so-called inverse coupling and $\xi$ is the bare anisotropic factor defined as the ratio of spatial lattice spacing to temporal lattice spacing $(a_s/a_t)$. Furthermore, $W_c={1 \over 3}\sum_c \Re e\  \Tr [1-U_c]$ where $U_c$ are the closed loops shown in Fig .  Tadpole improvement factors   $u_s = \langle {1 \over 3}\Re e\  \Tr [U_{s,s'}] \rangle^{1/4}$ and $u_t= \langle {1 \over 3} \Re e\ \Tr [U_{s,t}] \rangle^{1/2}/u_s$. This action has a smaller discretization error than the standard Wilson actions. Anisotropic actions $(\xi>1)$ have more time slices in plateaux compared to isotropic ones ($\xi=1$) as well. As a result, we obtain a better estimation for the effective mass \cite{Morningstar:1996ze}. In Table. \eqref{table:configuration_properties}, we list the properties of pure $SU(3)$ gauge configurations	used in this work.
\begin{figure}[htb]  	
	\centering
	\includegraphics[scale=1.2]{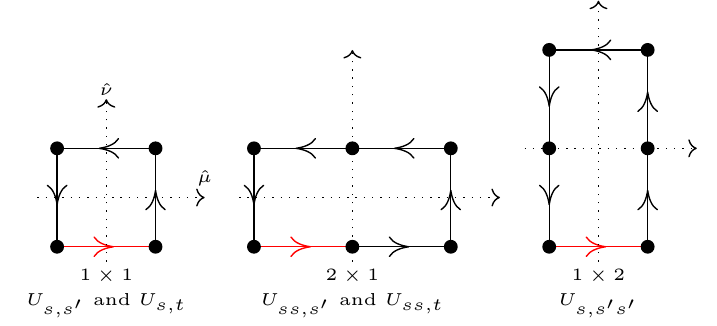}
	\caption{Closed loops $U_c$ included in the anisotropic tadpole improved action of Eq. \eqref{eq:S2}. The red links are updated in each step of the Monte Carlo method. \label{fig:improved_action}}
\end{figure}

\begin{table*}
	\begin{ruledtabular}
		\begin{tabular}{cccccccc|c|c}
		
			$\beta$&$\xi$&$\xi_R$&volume& $u_s$ &$u_t$& $a_s\sqrt{\sigma}$&$a_t\sqrt{\sigma}$&Smearing (space, time)& No. of configs\\
			\colrule
			4&4&3.6266(32)&$24\times96$&0.82006&1&0.3043(3)&0.0839(1)&
			$Stout_{0.15}^{20}$ ,Multihit(100)& 1060\\
			\colrule
		\end{tabular}
	\end{ruledtabular}
	\caption{Properties of the pure $SU(3)$ gauge configurations generated by the action of Eq. \eqref{eq:S2}. $\xi_R$ is the renormalized anisotropic factor computed based on the prescription described in Refs. \cite{Bicudo:2021tsc, Drummond:2002yg, Drummond:2003qu}. We used $100$ iterations of multihit on temporal gauge links, and $20$ iterations of stout with $\alpha=0.15$ on spatial gauge links.\label{table:configuration_properties}}
\end{table*}

\subsection{Smearing techniques}
The plateaux in the plots of effective mass, Eq. \eqref{eq:effective_mass},  for each charge distance $r$  usually appear in large time values where contamination from excited states are suppressed. However; short
distance fluctuations of the simulated gauge field are violent, therefore, as we compute the large
Wilson loops, the errors are accumulated and kill the signal of large distances. As a result, effective mass plots for large $t$ are very noisy. In addition
the error of the Monte Carlo method is proportional to $1/\sqrt{N}$, so it is costly to increase $N$,
the number of configurations, to compensate for the errors as it needs a large amount SU(3)
matrix manipulations.

To circumvent this problem, we apply two well known smearing techniques stout \cite{Morningstar:2003gk} and multihit  \cite{Brower:1981vt, Parisi:1983hm} smearing for spatial and temporal gauge links, respectively.
The multihit technique,   replaces each temporal link by its thermal average,
\begin{equation}
	U_4\rightarrow \bar{U}_4=\frac{\int dU_4 U_4 \,e^{\beta\Tr \left[U_4 F^\dagger\right]}}{\int dU_4 \,e^{\beta\Tr\left[ U_4 F^\dagger\right]}} \ .
\end{equation}
where $F$ is  part of the action connected to the temporal link $U_4$. Here, it is not possible to utilize the extended multihit technique as defined in Ref.  \cite{Cardoso:2013lla}, because our operators in the spatial Wilson line have a broader structure.

We should first tune the parameters of smearing procedures to increase the signal-to-noise ratio as much as is possible before trying to select the best operators. In this work, we skip this step and use the smearing parameters that we already tuned in \cite{Bicudo:2021tsc} to compute $\Sigma_g^+$ spectra. 
In particular, we use multihit with 100 iterations in time followed by stout smearing  in space with $\alpha=0.15$ and 20 iterations.

\subsection{Sieving the operators for each symmetry}
There are many different operators with a specific symmetry, but it is beneficial to choose operators which are close to the physical state of the system,  are easier to compute in term of time, and lead to a smaller statistical error in the correlation matrix and its diagonalization. For example, they include a smaller number of gauge links  while they lead a smaller amount of energy for the ground state of each symmetry. Furthermore, as our goal is to study a very excited flux tube, we prefer  to choose operators that do not lead to degeneracy in the spectra and have more excited states with a clear signal. To select the best operators, we describe the procedure for two case studies $\Sigma_u^-$ and $\Delta_g$ which are more illustrative. 
\begin{figure}[!th]
	\begin{minipage}{\columnwidth}
		\includegraphics[scale=0.8]{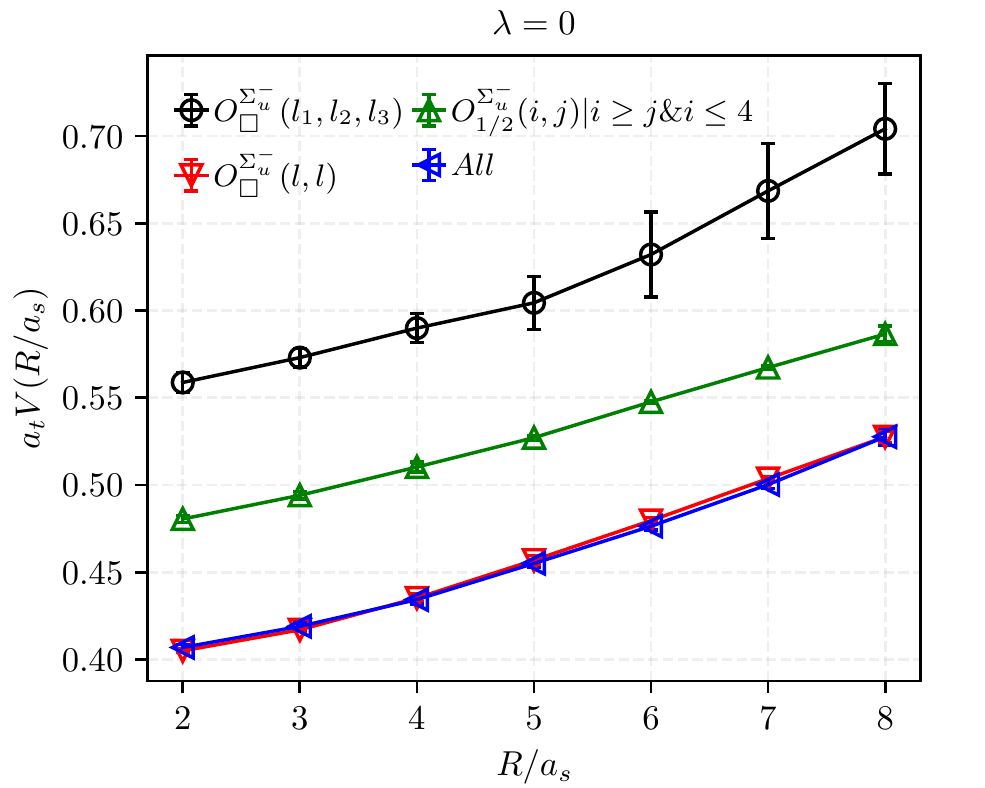}
		\subcaption{}
	\end{minipage}\vfil
	\caption{Selection of the operators by considering their overlap with the ground state of $\Sigma_u^-$, the red line and the blue one have more overlap because they lead to the lower energy for the ground state.}\label{fig:operator_selection} 
\end{figure}

In Fig. \eqref{fig:operator_selection} , we show the spectrum obtained using  $O_{\square}^{\Sigma_u^{-}}(l_1,l_2,l_3)$, $O^{\Sigma_u^{-}}_{1/2}(l_1, l_2)$, $O_{\square}^{\Sigma_u^{-}}(l,l)$. For now,  consider these operators just as operators with $\Sigma_u^-$ symmetry; their explicit formula does not matter for our goal in this section. By comparing the spectra,   we conclude that $O_{\square}^{\Sigma_u^{-}}(l_1,l_2,l_3)$ and $O^{\Sigma_u^{-}}_{1/2}(l_1, l_2)$ do not create a trial state as good as $O_{\square}^{\Sigma_u^{-}}(l,l)$ or their union. 
 When two types of operators lead to the same ground state approximately, we consider the excited states and choose the operator for which the excitations have better signal and the operator which is easier to compute.

\begin{figure}[!ht]
 	\begin{minipage}{\columnwidth}
 		\includegraphics[scale=0.8]{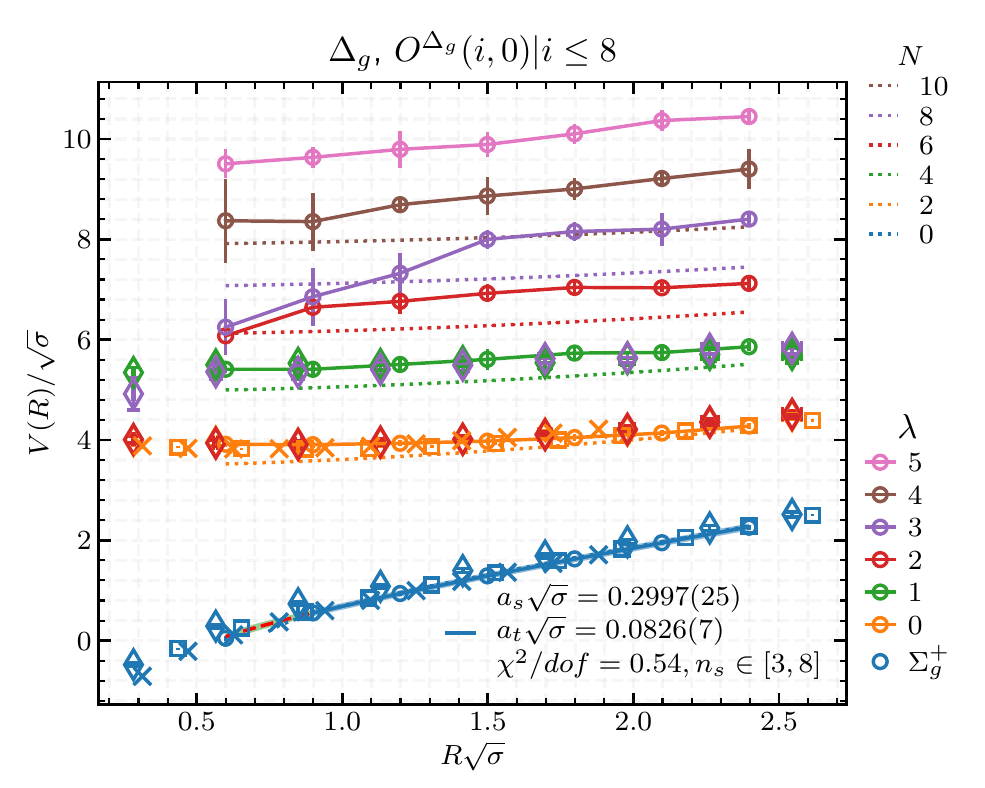}
 		\subcaption{}
 	\end{minipage}\vfil
 	\begin{minipage}{\columnwidth}
 		\includegraphics[scale=0.8]{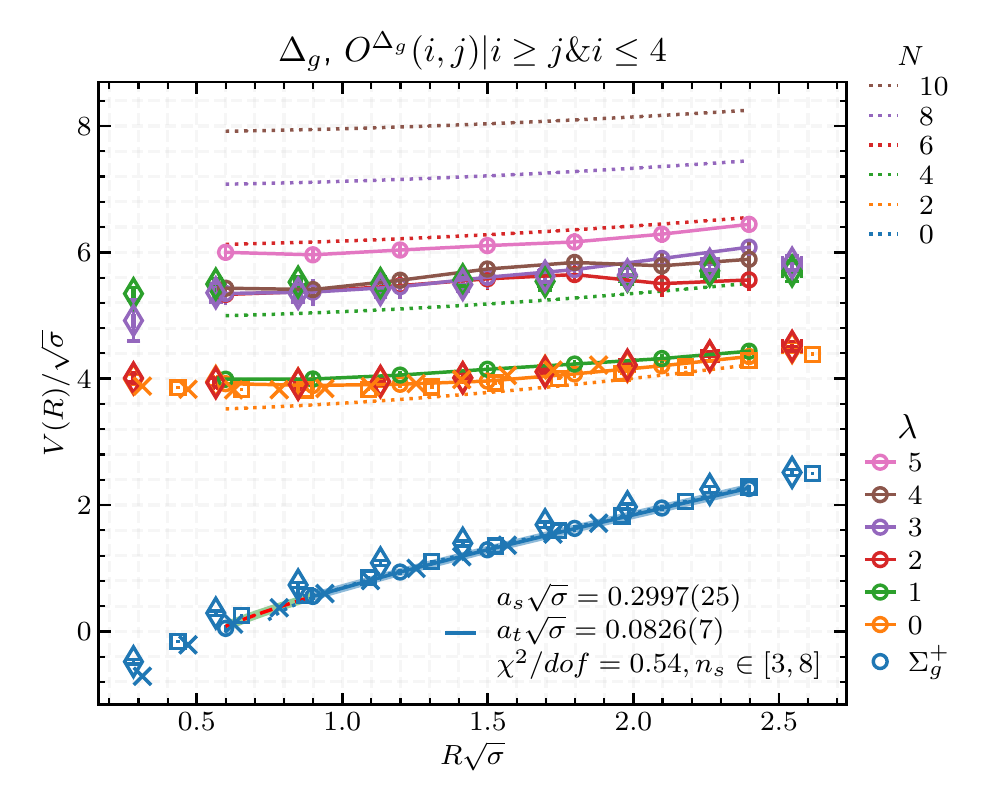}
 		\subcaption{}\label{fig:remove_degeneracy_b}
 	\end{minipage}
 	\caption{Operators $O^{\Delta_g}(i,0)$  used in (a) lead to smaller degeneracy than operator $O^{\Delta_g}(i,j)$ used in (b). Note that in (b), we use two class of operators together with $\epsilon=\pm1$ under $\mathcal{P}_x$. The diamond, square, and cross symbols show the data of  \cite{Morningstar:website1, Juge:1999ie,Juge:2002br}, \cite{Capitani:2018rox}, and \cite{Bicudo:2018jbb}, respectively. $\lambda=0$, $1$, $\ldots$ show the ground state, first excited state, etc., and $N$ indicates the corresponding Nambu-Goto level. $n_s \in [3,8]$ denotes the interval of the fit to the ground state to set the scale.} \label{fig:remove_degeneracy}
 \end{figure}

 When the operators pass the above criterion, to avoid the degeneracies in the spectra, we plot the spectra of each subset of operators or their unions and then select the subset that leads to smaller degeneracy. For example, in Fig. \eqref{fig:remove_degeneracy}, we prefer $O^{\Delta_g}(0,j)$ rather than $O^{\Delta_g}(1,j)$ with $\epsilon=\pm1$ under $\mathcal{P}_x$, because the obtained spectra by $O^{\Delta_g}(0,j)$ has no degeneracy. Moreover, for the sake of simplicity, computation of $O^{\Delta_g}(0,j)$ is faster and leads to a smaller systematic error as it includes a less number of operations on $SU(3)$ matrices. 
 
  In this step, we have used an ensemble of $50$  lattice configurations with volume $16^3\times64$ generated by the action of Eq. \eqref{eq:S2}, because the computation with lattice size is faster, hence, we can study a large numbers of operator (for example 80 operators) for some symmetries.
 
 As the overall conclusion for sieving the operators, when we use two or more types of operators, the excited states start to be degenerate while, in general, the ground state of each symmetry channel is improved. Furthermore, we  checked the consistency of our results with other results in the literature \cite{Bicudo:2018jbb,Capitani:2018rox, Morningstar:website1}. In Fig. \eqref{fig:remove_degeneracy}, we show two samples of this comparison.
 
 When the suitable class of operators is filtered out, we change the arbitrary parameter of the operators to cover the maximum possible length from the charge axis. In this way, we can make a tower of operators with the same symmetry but different distances from the charge axis. As the lattice configurations have periodic boundary conditions, we can select the parameters so that the set of operators sweeps the half length the lattice. 
  \subsection{Efficiency of the code}
  
  In this work, we use the public code \cite{sun:code} developed by our lattice QCD group to use graphics processing unit (GPU) for lattice QCD computation in quenched approximation. The code has been written using CUDA C. It is able to generate $SU(N)$  pure gauge lattice   configurations \cite{Cardoso:2011xu} based on the Wilson isotropic and anisotropic action, tadpole improved action. It can apply different types of well known smearing techniques such as multihit \cite{Brower:1981vt, Parisi:1983hm}, extended-multihit \cite{Cardoso:2013lla}, APE \cite{Albanese:1987ds}, stout \cite{Morningstar:2003gk}, HYP \cite{Hasenfratz:2001hp}. Also, it does gauge fixing \cite{Cardoso:2012kbk,Cardoso:2014una,Cardoso:2012uu}, computes the Wilson loop, flux tube profile \cite{Bicudo:2018jbb}, and it is extensible for doing new tasks. In Table. \eqref{table:code_efficiency}, we summarize the time needed for different parts of our computations executed by a GeForce RTX 2080 Ti with 7.5 cc\footnote{Compute Capability} in single precision.
  \begin{table}[!h]
  	\begin{ruledtabular}
  		\begin{tabular}{p{0.35\linewidth}|p{0.15\linewidth} | p{0.45\linewidth}}
  			
  			Task& Time (s)&Details\\
  			\colrule
  			Generating a  $SU(3)$ lattice configuration with the volume $24^3\times96$ & $3$ & $50$ iteration, each iterations includes $4$ heat bath and $7$ over-relaxation steps. \\
  			\hline
  			Stout smearing&$0.014$ & $20$ iterations\\
  			\hline
  			Multihit smearing&$0.8$& $100$ iterations\\
  			\hline
  			Computation of the correlation matrix & $350$ & $12\times12$ matrix, the time depends on the operators.
  		\end{tabular}
  	\end{ruledtabular}
  	\caption{The time takes for different tasks executed by a GeForce RTX 2080 Ti with 7.5 cc in single precision for the tadpole improved action Eq. \eqref{eq:S2}. \label{table:code_efficiency}}
  \end{table}
  
\section{Flux tubes spectra\label{sec:spectra}}
 In this section, we elaborate more details about the selected operator based on the criteria of the last section. Some suboperators were filtered out for different symmetries. e show suboperators and their explicit formula for different symmetries obtained based on the prescription of Sec. \ref{sec:operators_construction}. Besides, we show the spectra of each related flux at the same time.
 
 \subsection{Analysis of our results}
 In the following sections, we show the operators and obtained spectra  for different symmetries of the flux tube, up to $\Delta_u$, as a function of the charge distance $R$ .To set the scale, we fit the ground state of $\Sigma_g^+$ to $V(r)=V_0+\sigma r$ \cite{Bicudo:2021tsc}. Then,  we use the value of $\sigma$ to present our results in the string tension unit ($\sqrt{\sigma}$).  The value of $\sigma\approx 0.18\ GeV^2$ \cite{Godfrey:1985xj}.
 
 To set the scale for anisotropic actions, we also need the value of renormalized anisotropy $\xi_R=a_s/a_t $. The renormalized anisotropy $\xi_R$ can be determined as a function of the bare anisotropy $\xi$. The ground state potential is computed with the Wilson loop by considering two different directions for the time direction, once with the anisotropic direction and once with one of the isotropic directions. Then, by comparing the short distance potential for both time directions, the ratio $a_s/a_t$ is determined. 
We follow exactly the same procedure as in Ref. \cite{Bicudo:2021tsc} where all the details are presented based on Refs.\cite{Drummond:2002yg, Drummond:2003qu}.

 When we compare our result with the Nambu-Goto spectrum, Eq. \eqref{eq:nambu_goto}, we observe excited states departure from the Nambu-Goto model. To quantify this tension, we fit our results to
 \begin{align}
 	&V_1(R)=\sigma R\sqrt{1+\frac{2\pi}{\sigma_2 R^2} (N-(D-2)/24)},
 	\label{eq:modified_nambu_goto}
 \end{align}
 where $\sigma_2< 1$ corresponds to a larger gap between excited states than in the Nambu-Goto string model.  
 
 We fit the data of all excitations with clear signals to Eq. \eqref{eq:modified_nambu_goto}. Notice that the corresponding excitation numbers $N$ are fixed and we exclude the data for small charge separations $r$.  \textcolor{black}{The values of $N$ for each symmetry are shown in Table. \eqref{table:quantum_numbers}.}  
 \iffalse
 The notations in Fig. \eqref{fig:all_spectra} are as the following:
 \begin{itemize}
 	\item Dashed lines show the spectrum of the modified Nambu-Goto ansatz Eq. \eqref{eq:modified_nambu_goto}, and dotted lines show the spectrum of the Nambu-Goto model Eq. \eqref{eq:nambu_goto}.
 	\item The width of highlighted areas around dashed lines shows the error of the fits.
 	\item The vertical gray highlighted areas show the interval of charge separations $r$ included in the fit.
 	\item $\lambda=0,1,\ldots$ show the ground state, first excited, and so on. For each $\lambda$, the corresponding Nambu-Goto level $N$ has the same color as the data.
 	\item Fitted for $N_i\to N_f$ in the title of figures denote to the interval of $N$ included in the multifit to Eq. \eqref{eq:modified_nambu_goto}.
 	\item We did not fit $\Sigma_u^-$ spectrum to Eq. \eqref{eq:modified_nambu_goto} because the slope of states are steeper than the corresponding  Nambu-Goto level, the dotted lines in Fig. \eqref{fig:sigma_um_spectra}. \textcolor{violet}{ Furthermore, we exclude the ground state data of $\Sigma_g^-$ to be able to fit Eq. \eqref{eq:modified_nambu_goto} to the rest of the data. }
 \end{itemize}
 \fi
 \subsection{Operators for $\Sigma_g^+$, $\Pi_u$, $\Delta_g$ symmetries and the spectra}
\begin{figure*}[ht]
 	\begin{minipage}{\columnwidth}
 		\includegraphics[scale=0.9]{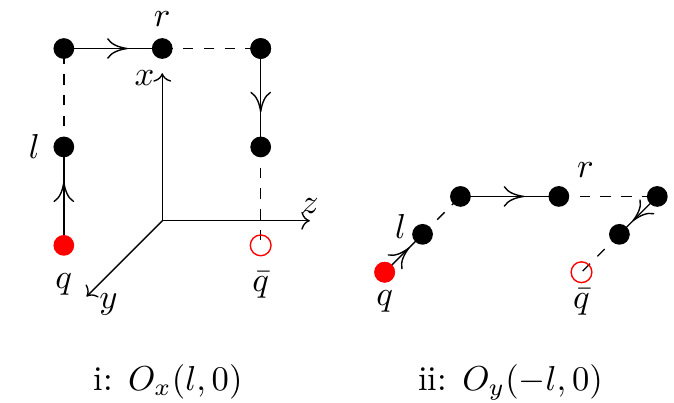}
 		\subcaption{We vary $l$  from 0 (for $\Sigma_g^+$) or 1 to 12 to construct 13 or 12 operators. In $O_x(l,0)$, $0$ is to emphasize the suboperators are planer.  \label{op:Oi}}
 	\end{minipage}\hfil
 	\begin{minipage}{\columnwidth}
 		\includegraphics[scale=0.8]{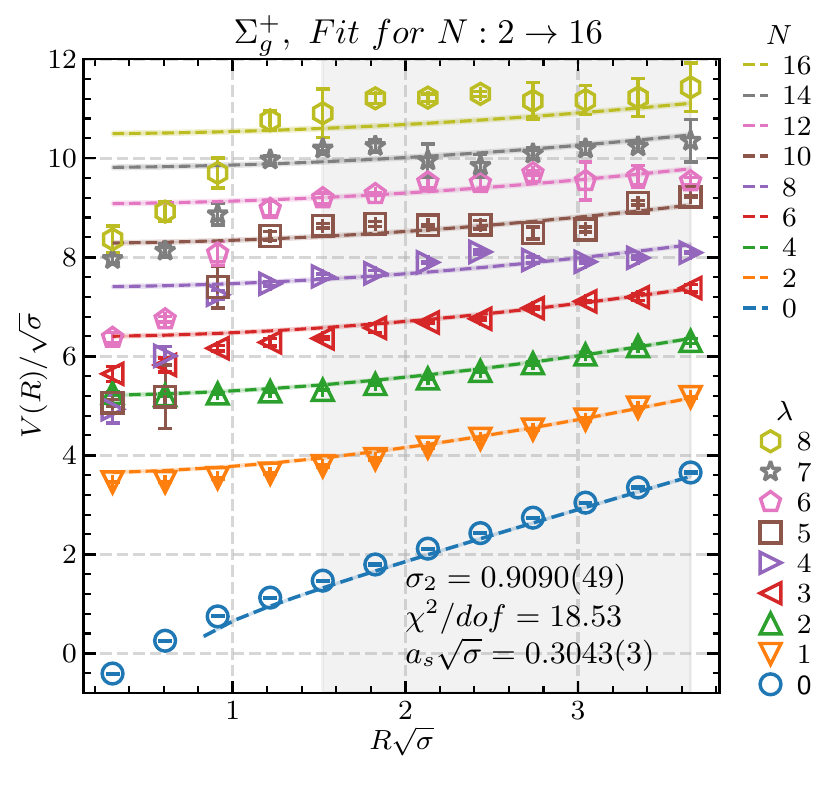}
 		\subcaption{$\Sigma_g^+$ spectrum, data derived from \cite{Bicudo:2021tsc}.\label{fig:sigma_gp_spectra}}
 	\end{minipage}\vfil	
 	\begin{minipage}{\columnwidth}
 		\includegraphics[scale=0.8]{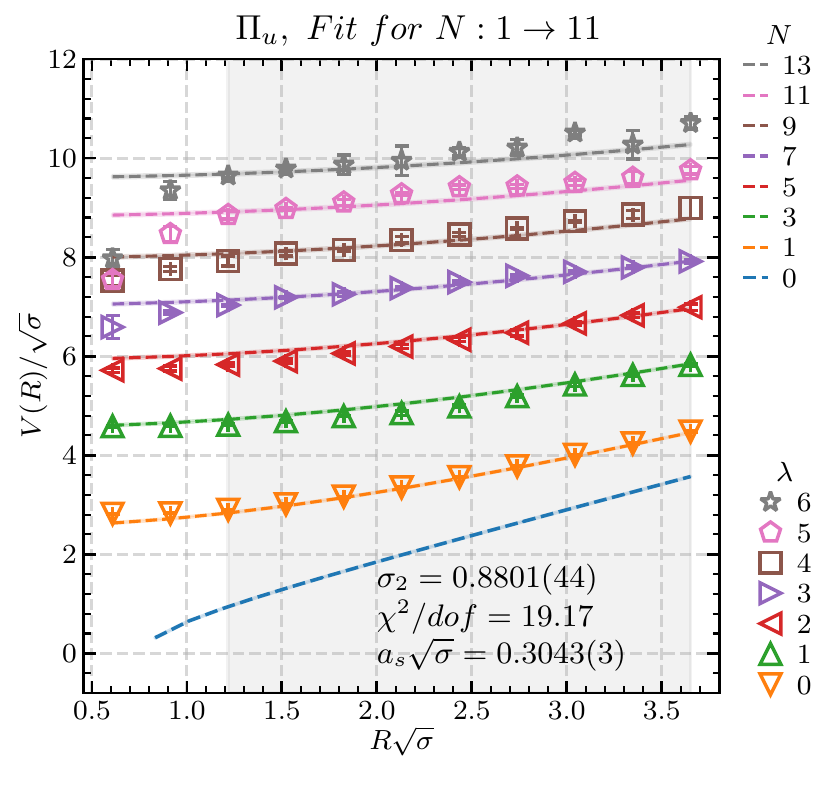}
 		\subcaption{$\Pi_u$ spectrum.\label{fig:pi_u_spectra}}
 	\end{minipage}\hfil
 	\begin{minipage}{\columnwidth}
 		\includegraphics[scale=0.8]{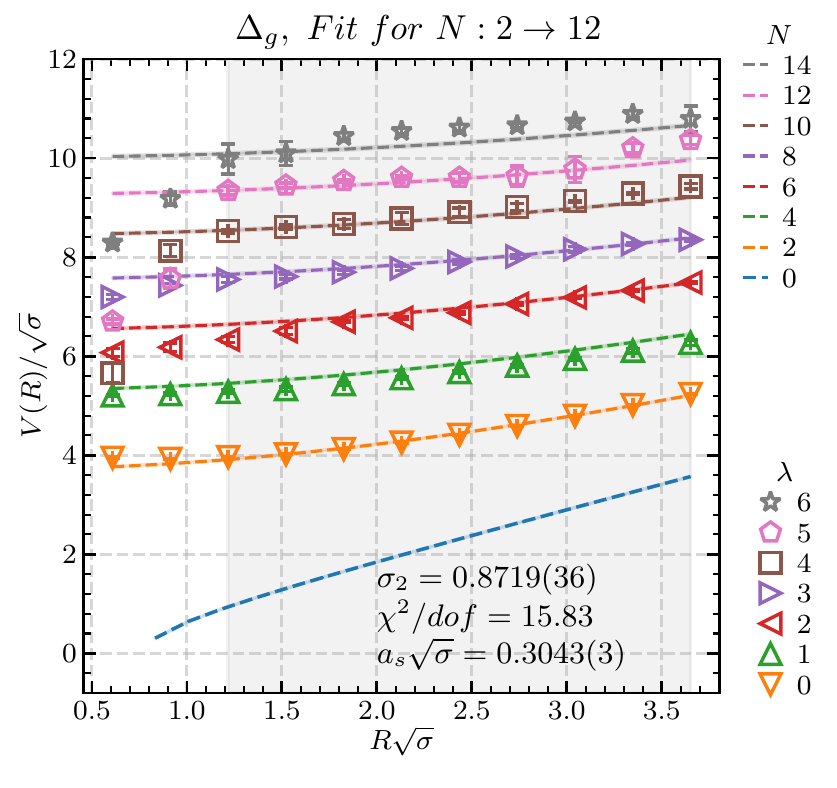}
 		\subcaption{$\Delta_g$ spectrum.\label{fig:delta_g_spectra}}
 	\end{minipage}
 	\caption{$\Sigma_g^+$, $\Pi_u$, and $\Delta_g$ spectra. Dashed \spectraCaption }
 \end{figure*}

 We use suboperators shown in Fig. \eqref{op:Oi} to construct operator with $\Sigma_g^+$, $\Pi_u$, and $\Delta_g$ symmetries. Consequently, the operators are written as
 \begin{eqnarray}
 	O^{\Sigma_g^+}(l, 0)=\frac{1}{2}&\bigg(&O_x(l,0)+O_y(l,0)\nonumber\\&&+O_x(-l,0)+O_y(-l, 0)\bigg),\\
 	O^{\Pi_u}(l, 0)=\frac{1}{2}&\bigg(&O_x(l,0)+ \im O_y(l,0)\nonumber\\&&-O_x(-l,0)-\im O_y(-l, 0)\bigg),\\
 	O^{\Delta_g}(l, 0)=\frac{1}{2}&\bigg(&O_x(l,0)-O_y(l,0)\nonumber\\&&+O_x(-l,0)-O_y(-l, 0)\bigg).
 \end{eqnarray}
 The spectra obtained using these operators are shown in Figs. \eqref{fig:sigma_gp_spectra}, \eqref{fig:pi_u_spectra}, and \eqref{fig:delta_g_spectra}, respectively.

 \begin{figure*}[!th]
 	\begin{minipage}{\columnwidth}
 		\centering
 		\includegraphics[scale=0.85]{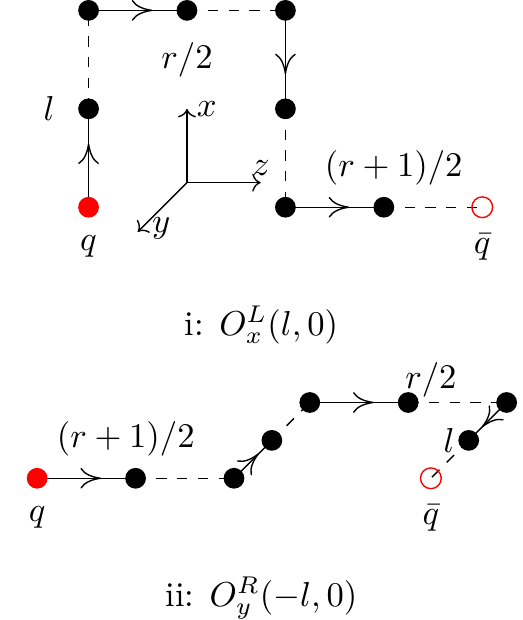}
 		\subcaption{We vary $l$ from $1$ to $12$ to build $12$ operators.\label{op:halfline}}
 	\end{minipage}\hfil
 	\begin{minipage}{\columnwidth}
 		\includegraphics[scale=0.8]{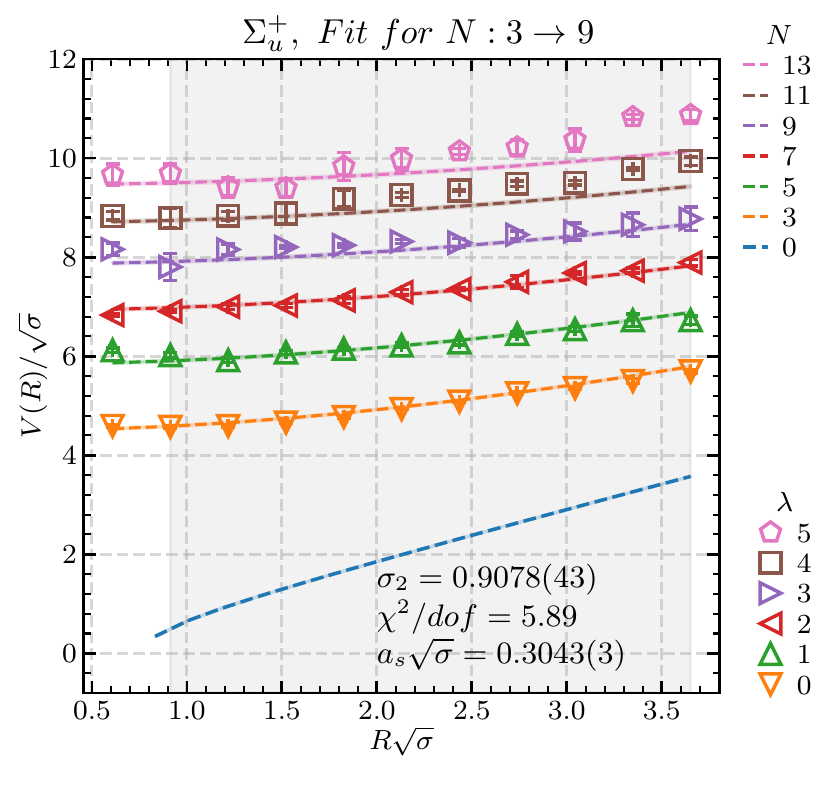}
 		\subcaption{$\Sigma_u^+$ spectrum \label{fig:sigma_up_spectra}}
 	\end{minipage}\vfil
 	\begin{minipage}{\columnwidth}
 		\includegraphics[scale=0.8]{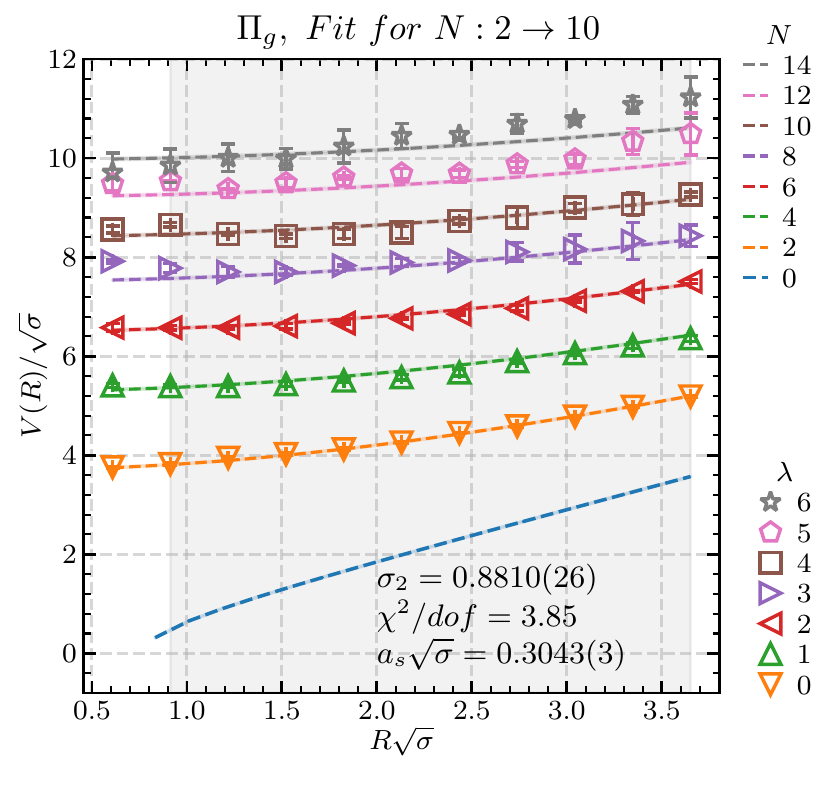}
 		\subcaption{$\Pi_g$ spectrum.\label{fig:pi_g_spectra}}
 	\end{minipage}\hfil
 	\begin{minipage}{\columnwidth}
 		\includegraphics[scale=0.8]{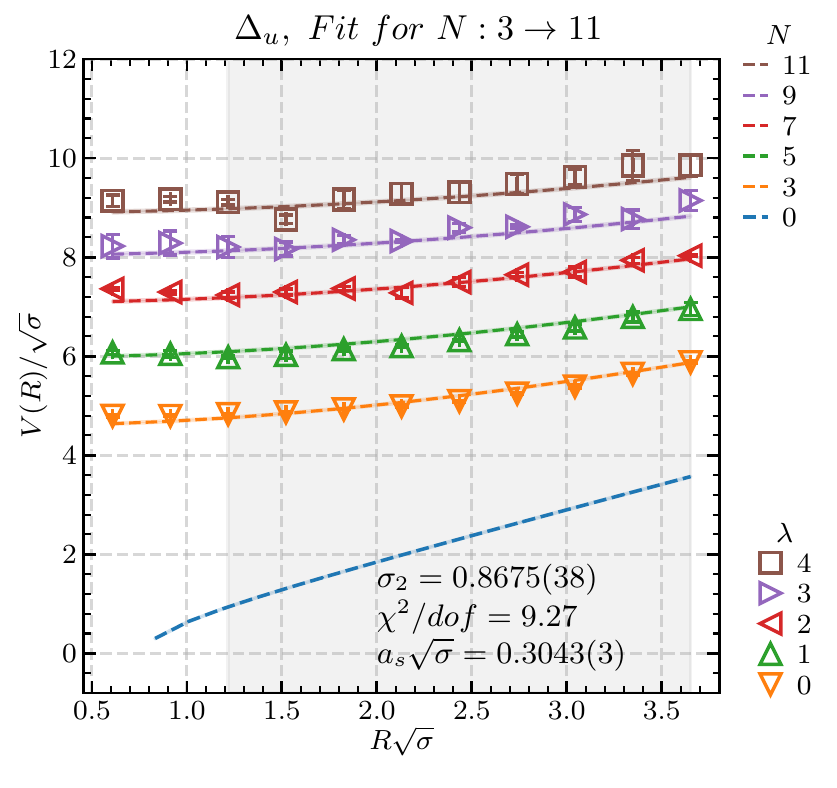}
 		\subcaption{$\Delta_u$ spectrum.\label{fig:delta_u_spectra}}
 	\end{minipage}
 	\caption{$\Sigma_u^+$, $\Pi_g$, and $\Delta_u$ spectra. Dashed \spectraCaption}
 \end{figure*}
 \subsection{Operators for $\Sigma_u$, $\Pi_g$, $\Delta_u$ symmetries and the spectra}
 For $\Sigma_u^+$, $\Pi_g$, $\Delta_u$  symmetries of the flux tube, we use half-line suboperator shown in Fig. \eqref{op:halfline}. The explicit formulas of these symmetries are obtained as
 \begin{eqnarray}
 	O^{\Sigma_u^+}_{1/2}(l,0)=\frac{1}{2\sqrt{2}}&\bigg(&O^L_x(l,0)+O^L_y(l,0)+O^L_x(-l,0)
 	\nonumber\\&&+O^L_y(-l,0)-O_x^R(-l, 0)-O^R_y(-l,0)\nonumber\\
 	&&-O_x^R(l, 0)-O_y^R(l, 0)\bigg),
 \end{eqnarray}
\begin{eqnarray}
 	O^{\Pi_g}_{1/2}(l,0)=\frac{1}{2\sqrt{2}}&\bigg(&
 	O_x^L(l,0)+\im O_y^L(l,0)-O_x^L(-l,0)\nonumber\\&&-\im O_y^L(-l,0)-\big[ O_x^R(l,0)+\im O_y^R(l,0)
 	\nonumber\\&&-O_x^R(-l,0)-\im O_y^R(-l,0)\big]
 	\bigg),
 \end{eqnarray}
  \begin{eqnarray}
 	O^{\Delta_u}_{1/2}(l,0)=\frac{1}{2\sqrt{2}}&\bigg(&
 	O_x^L(l,0)- O_y^L(l,0)+O_x^L(-l,0)\nonumber\\
 	&&- O_y^L(-l,0)-\big[ O_x^R(l,0)- O_y^L(l,0)\nonumber\\
 	&&+O_x^R(-l,0)- O_y^R(-l,0)\big]
 	\bigg). 
 \end{eqnarray}
We use $1/2$ in the indices of these operator to denote they are made of half line suboperators.
 In Figs. \eqref{fig:sigma_up_spectra}, \eqref{fig:pi_g_spectra}, and \eqref{fig:delta_u_spectra}, we show the spectra obtained using these operators.
 \subsection{Operators for $\Sigma_g^-$ symmetry and the spectrum}
\begin{figure*}[!ht]
 	\begin{minipage}{\columnwidth}
 		\centering
 		\includegraphics[scale=0.9]{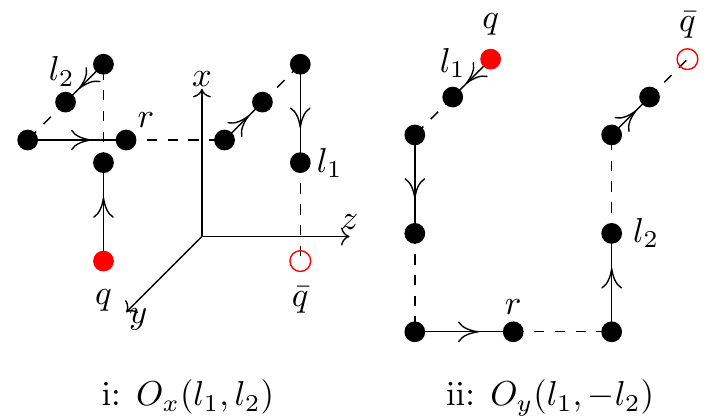}
 		\subcaption{We fix $l_1=1$ and let $l_2$ vary from 1 to 12, so we end up with 12 operators.\label{fig:sigma_gm_operator}}
 	\end{minipage}\hfil
 	\begin{minipage}{\columnwidth}
 		\centering
 		\includegraphics[scale=0.8]{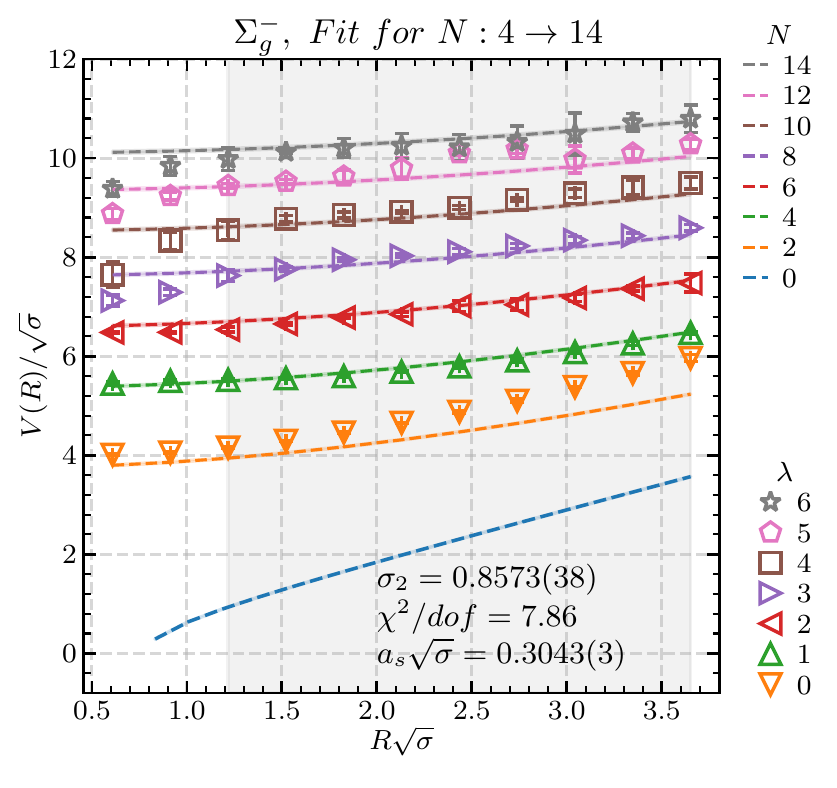}
 		\subcaption{$\Sigma_g^-$ spectrum. \label{fig:sigma_gm_spectra}}
 	\end{minipage}
 	\caption{ $\Sigma_g^-$ suboperators and the spectrum. Dashed \spectraCaption}
 \end{figure*}

 To construct operators with $\Sigma_g^-$ symmetry, we use the suboperators shown in Fig. \eqref{fig:sigma_gm_operator}. There are several options for the values of $l_1$ and $l_2$ in Eq. \eqref{eq:sigma_gm_operator}. We choose values for $l_2$ that range from 1 to 12 while fixing the value of $l_1=1$,
 \begin{eqnarray}
 	O^{\Sigma_g^-}(l_1, l_2)=\frac{1}{2\sqrt{2}}&&\bigg(O_x(l_1,l_2)-O_x(l_1,-l_2)
 	\nonumber\\&&+O_x(-l_1,-l_2)-O_x(-l_1,l_2)\nonumber\\
 	&&+O_y(l_1,-l_2)-O_y(-l_1,-l_2)
 	\nonumber\\
 	&&+O_y(-l_1,l_2)-O_y(l_1,l_2)\bigg).\label{eq:sigma_gm_operator}
 \end{eqnarray}

\begin{figure*}[!ht]
	\begin{minipage}{\columnwidth}
		\centering
		\includegraphics[scale=0.7]{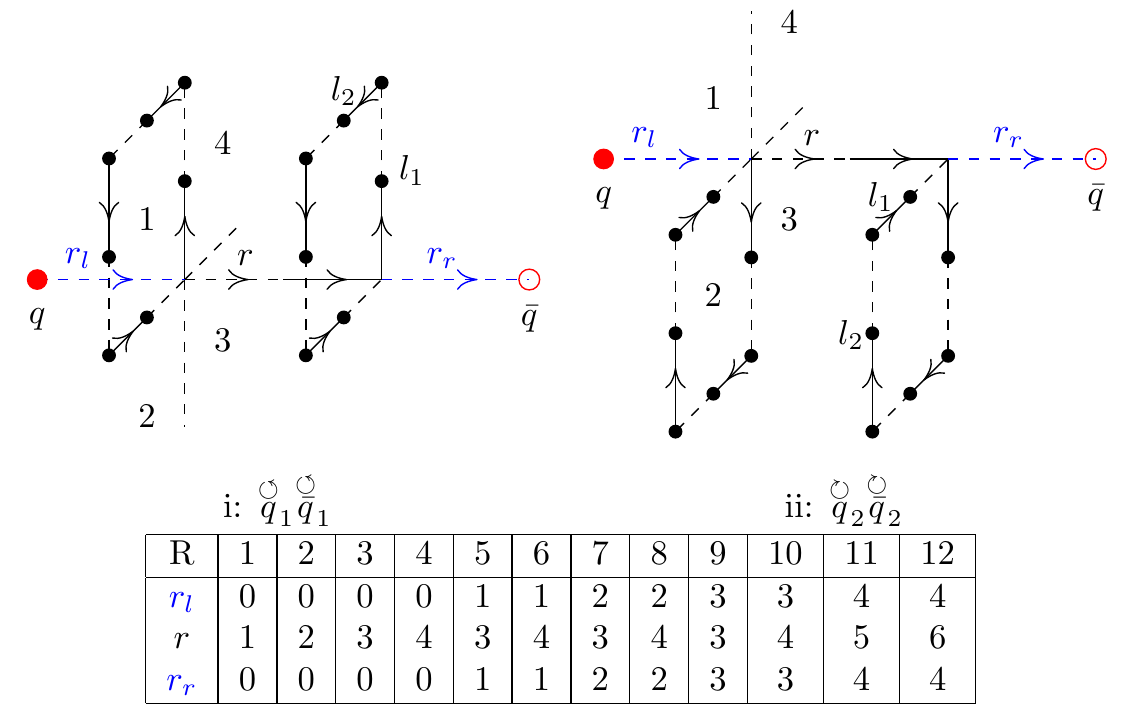}
		\subcaption{We select $l_1=l_2$ and let $l_1$ vary from 1 to 12 to build 12 operators.  The values of $r_l$ and $r_r$ derived from Ref. \cite{Capitani:2018rox}. \label{fig:op_sigma_um}}
	\end{minipage}\hfil
	\begin{minipage}{\columnwidth}
		\centering
		\includegraphics[scale=0.8]{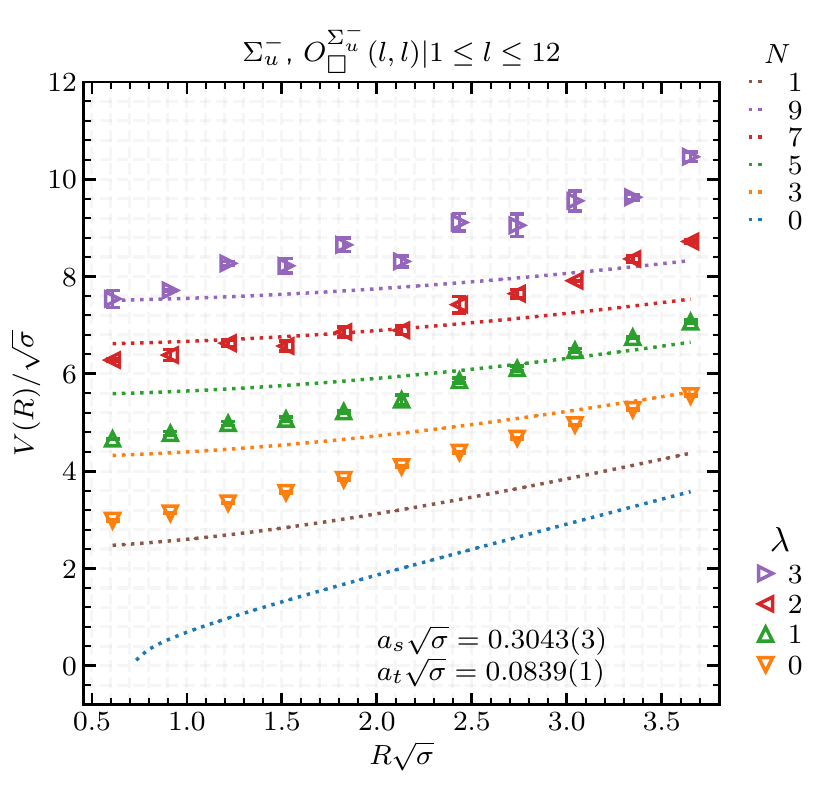}
		\subcaption{$\Sigma_u^-$ spectrum. \label{fig:sigma_um_spectra}}
	\end{minipage}
	\caption{$\Sigma_u^-$  suboperators and the spectrum, dotted lines show the Nambu-Goto model spectrum, Eq. \eqref{eq:nambu_goto}. $\lambda=0,\ 1,\ 2, \ldots$ denote the ground state, first excitation, and so on,  depicted with the same color as the corresponding quantum number $N$.}
\end{figure*}
 
 In Fig. \eqref{fig:sigma_gm_spectra}, we could  fit  the ansatz of Eq. \eqref{eq:modified_nambu_goto} to our data properly only if we exclude the ground state data (orange points) from the fit and assign $N=4$ to the first excitation (green points). This is in contrast to the prediction of string models  which assign $N=4$ to the ground state of $\Sigma_g^-$ \cite{Juge:2003ge}. Therefore, this state might correspond to another particle, say an axion \cite{Dubovsky:2013gi,Athenodorou:2021vkw}. 
 %	What supports this idea is the energy of this state which locates almost $1.85\sqrt{\sigma}$ higher than the flux tube ground state ($\Sigma_g^+$) which is consistent with the ground state of axions \cite{Dubovsky:2013gi,Athenodorou:2021vkw}.} 

\subsection{The operator for $\Sigma_u^-$ symmetry and the spectrum.}
For $\Sigma_u^-$ symmetry, we use loop like suboperators shown in Fig. \eqref{fig:op_sigma_um}. So we obtain 
\begin{eqnarray}
O^{\Sigma_u^-}_{\square}(l,l)
=\frac{1}{2\sqrt{2}}&\bigg(&\qu{1}\aqu{1}+\qu{2}\aqu{2}+\qu{3}\aqu{3}
\nonumber\\&&+\qu{4}\aqu{4}-\quc{4}\aquc{4}-\quc{3}\aquc{3}\nonumber\\
&&
-\quc{2}\aquc{2}-\quc{1}\aquc{1}\bigg).
\end{eqnarray}
In Fig. \eqref{fig:sigma_um_spectra}, we show the $\Sigma_u^-$ spectrum. Note that the slope of the energy levels are steeper than the Nambu-Goto spectrum, so we did not fit the ansatz of Eq. \eqref{eq:modified_nambu_goto} to the data. 

\section{Discussion on $\Sigma_{g}^-$ and $\Sigma_{u}^-$ spectra}\label{sec:axion_glueball}
\begin{figure*}
\begin{minipage}{\columnwidth}
	\includegraphics[scale=0.8]{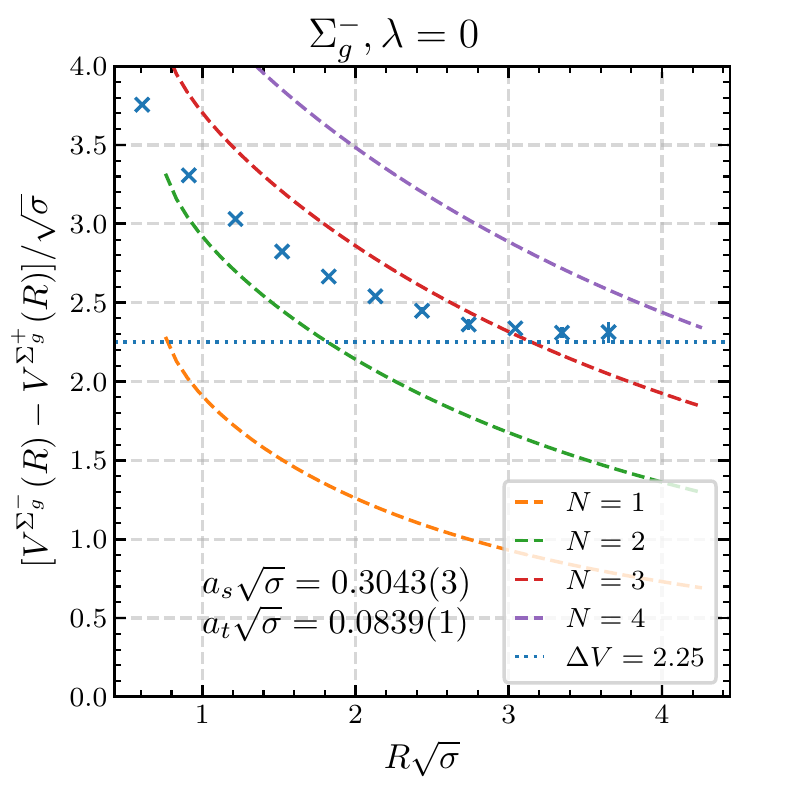}	
	\subcaption{}
\end{minipage}\hfil	
\begin{minipage}{\columnwidth}
	\includegraphics[scale=0.8]{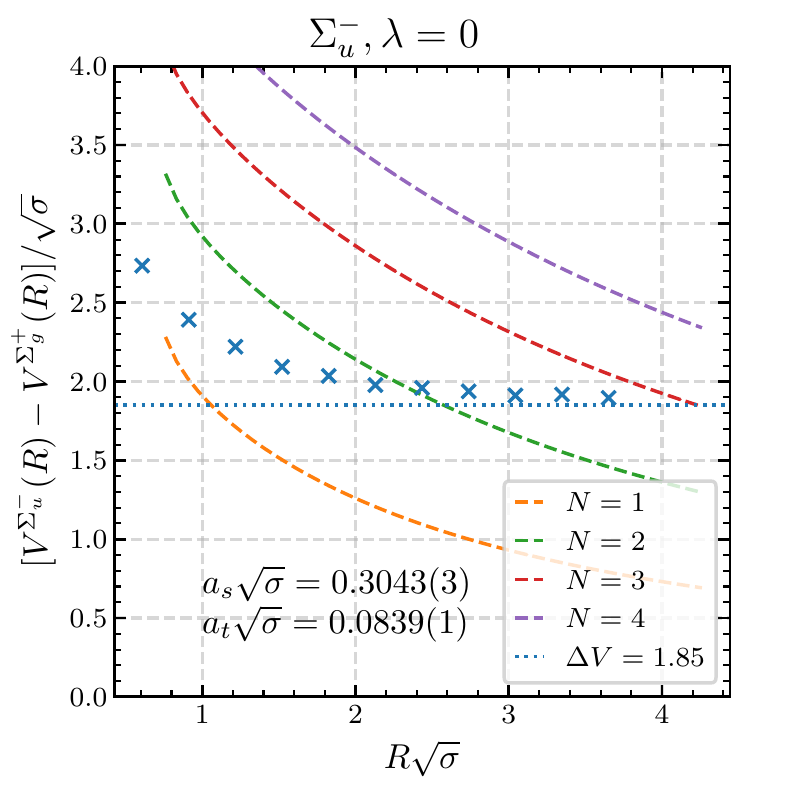}	
	\subcaption{}
\end{minipage}\vfil
\begin{minipage}{\columnwidth}
	\includegraphics[scale=0.8]{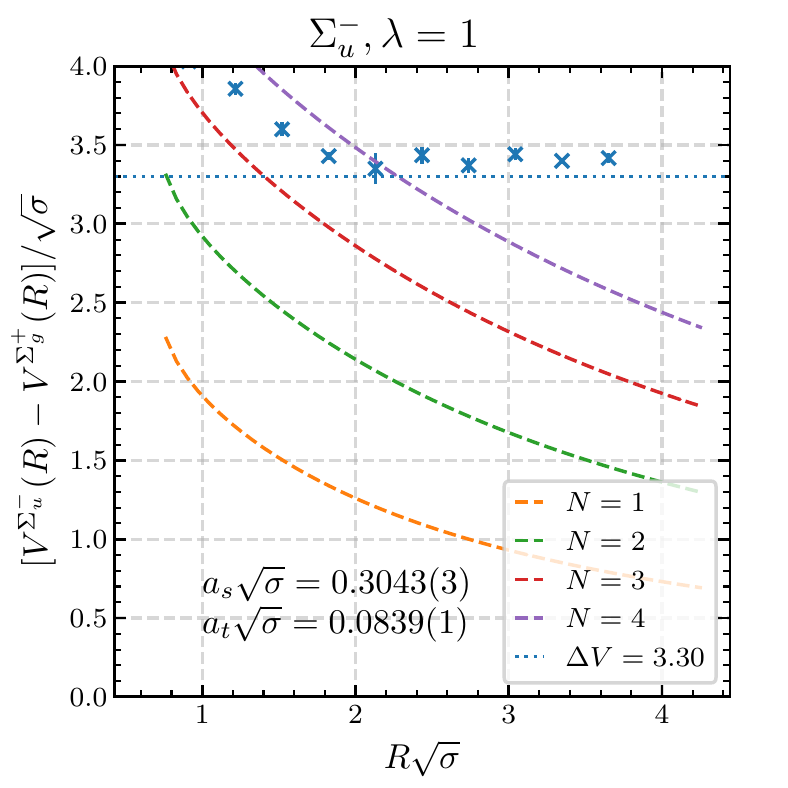}	
	\subcaption{}
\end{minipage}
\caption{Subtraction of $V_{\Sigma_g^+}(R)$ from the ground state $\lambda=0$ and the first excited state $\lambda=1$ of $\Sigma_g^-$ and $\Sigma_u^-$. Dashed lines show the subtraction  of the Nambu-Goto model ground state from the excitation N. The horizontal dotted lines show the approximate value of the plateau, note that it is not  fitted to the data.\label{fig:axion}}	
\end{figure*}

The ground state of $\Sigma_g^-$,  $\Sigma_u^-$ , and excited states of $\Sigma_u^-$ flux tube are inconsistent with the Nambu-Goto string model. This deviation might be due to the coupling of another particle, say an axion, to the flux tube. To find the mass of the coupled particle, we subtract the ground state of the flux tube, $\Sigma_g^+$, from these spectra, the result shown in Fig. \eqref{fig:axion}. It is interesting in Figs. (\eqref{fig:axion}. a-c) there are plateaux. The values obtained for the plateaux are $\Delta V_{{\Sigma_g^-}, \Sigma_g^+}/\sqrt{\sigma}\approx2.25$, $\Delta V_{{\Sigma_u^-}, \Sigma_g^+}/\sqrt{\sigma}\approx 1.85$ and $\Delta V_{{\Sigma_u^-}^*, \Sigma_g^+}/\sqrt{\sigma}\approx 3.30$. The value of $1.85\sqrt{\sigma}$ is already reported for the mass of an axion computed based on the spectra of the closed flux tube \cite{Andreas_2017}. Also, the mass of the lowest lying scalar glueball has been reported as  $M_{0^{++}}/\sqrt{\sigma}=3.607(87)$ \cite{Lucini_2001} which is a little bit higher than  $\Delta V_{{\Sigma_u^-}^*, \Sigma_g^+}/\sqrt{\sigma}$. 

\section{Conclusions and outlook \label{sec:conclusion}}
We succeeded to compute a significant number of excitations for different symmetries of the flux tube, improving the state of the art, Table. \eqref{Tab:number_excitations}. Notice that the key difference between the operators we used in this paper and the literature is that, our operators have the same spatial deformation but they sweep the width of the flux tube.

\begin{figure}[!h]
	\includegraphics[scale=1]{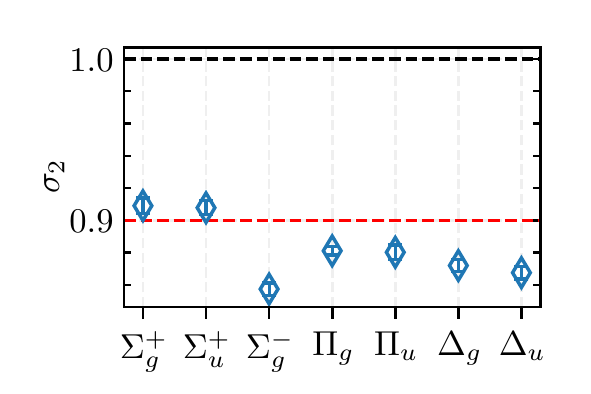}
	\caption{Values of $\sigma_2$, Eq. \eqref{eq:modified_nambu_goto}, for different symmetries.\label{fig:deviation_nambu}}
\end{figure}

Considering a second parameter $\sigma_2$ in the Arvis potential, Eq. \eqref{eq:modified_nambu_goto}, results in better fits to the data. The values of $\sigma_2$ are almost $10\%$ smaller than $\sigma$, Fig. 
\eqref{fig:deviation_nambu}, leading to larger energy splitting between energy levels than the Nambu-Goto spectrum. This tension may be a signal for the existence of a constituent gluon \cite{Mueller:2019mkh} in the excited flux tubes. As the $\chi^2/dof$ of the fits are large,  we should take this deviation with a grain of salt. Furthermore, this deviation also depends on the lattice artifacts, because it is smaller for the anisotropic Wilson action \cite{Bicudo:2021tsc}.

Furthermore, by analyzing  $\Sigma_g^-$ and $\Sigma_u^-$ spectra, we found the signal for the coupling of another particle to these flux tubes. Trying new types of operator for these symmetries, finding methods to increase the signal-to-noise ratio, and studying the flux tube profile \cite{Bicudo:2018jbb}, especially for $\Sigma_u^-$ which has a smaller energy,  will help to understand why they behave differently from the Nambu-Goto string model.

\begin{table}
\begin{ruledtabular}
	\begin{tabular}{c|c|c|c|c|c|c|c|c}
	
		{Action}{	$\Lambda_\eta^+$}&	$\Sigma_g^+$&$\Sigma_g^-$&$\Sigma_u^+$&$\Sigma_u^-$&$\Pi_g$&$\Pi_u$&$\Delta_g$&$\Delta_u$\\
\colrule
		$S_{II}$&2&0&0&0&1&1&1&0\\
		\colrule
&8&6&4&2&6&6&6&4
	\end{tabular}
\end{ruledtabular}
	\caption{The first row shows the numbers of excitations are already reported in the literature \cite{Juge:1999ie,Juge:2002br}, the second row show our results.\label{Tab:number_excitations}}
\end{table}

\begin{acknowledgments}
We acknowledge the discussion on flux tubes and our results with our colleagues Andreas Athenodorou, Bastian Brandt, Marco Cardoso, Kate Clark, Mika Lauk, Lasse Müller, Emílio Ribeiro, Caroline Riehl, and Marc Wagner.
Alireza Sharifian and Nuno Cardoso  are supported by FCT under the Contract No. SFRH/PD/BD/135189/2017 and SFRH/BPD/109443/2015, respectively. The authors thank CeFEMA, an IST research unit whose activities are partially funded by
FCT contract  UIDB/04540/2020 for R\&D Units. 
PB also is grateful to the
Pauli Institute for Theoretical Studies Visiting Researcher Program
and the hospitalities of the Institute of Nuclear Physics of the
Polish Academy of Sciences and of the Institute for Theoretical Physics of ETH Zurich.
\end{acknowledgments}

\bibliographystyle{apsrev4-2}

\bibliography{references}% Produces the bibliography via BibTeX.

\end{document}